\hsize=31pc 
\vsize=49pc 
\lineskip=0pt 
\parskip=0pt plus 1pt 
\hfuzz=1pt   
\vfuzz=2pt 
\pretolerance=2500 
\tolerance=5000 
\vbadness=5000 
\hbadness=5000 
\widowpenalty=500 
\clubpenalty=200 
\brokenpenalty=500 
\predisplaypenalty=200 
\voffset=-1pc 
\nopagenumbers      
\catcode`@=11 
\newif\ifams 
\amsfalse 
%
%
%
\newfam\bdifam 
\newfam\bsyfam 
\newfam\bssfam 
\newfam\msafam 
\newfam\msbfam 
\newif\ifxxpt    
\newif\ifxviipt  
\newif\ifxivpt   
\newif\ifxiipt   
\newif\ifxipt    
\newif\ifxpt     
\newif\ifixpt    
\newif\ifviiipt  
\newif\ifviipt   
\newif\ifvipt    
\newif\ifvpt     
%
%
\def\headsize#1#2{\def\headb@seline{#2}%
                \ifnum#1=20\def\HEAD{twenty}%
                           \def\smHEAD{twelve}%
                           \def\vsHEAD{nine}%
                           \ifxxpt\else\xdef\f@ntsize{\HEAD}%
                           \def\m@g{4}\def\s@ze{20.74}%
                           \loadheadfonts\xxpttrue\fi 
                           \ifxiipt\else\xdef\f@ntsize{\smHEAD}%
                           \def\m@g{1}\def\s@ze{12}%
                           \loadxiiptfonts\xiipttrue\fi 
                           \ifixpt\else\xdef\f@ntsize{\vsHEAD}%
                           \def\s@ze{9}%
                           \loadsmallfonts\ixpttrue\fi 
                      \else 
                \ifnum#1=17\def\HEAD{seventeen}%
                           \def\smHEAD{eleven}%
                           \def\vsHEAD{eight}%
                           \ifxviipt\else\xdef\f@ntsize{\HEAD}%
                           \def\m@g{3}\def\s@ze{17.28}%
                           \loadheadfonts\xviipttrue\fi 
                           \ifxipt\else\xdef\f@ntsize{\smHEAD}%
                           \loadxiptfonts\xipttrue\fi 
                           \ifviiipt\else\xdef\f@ntsize{\vsHEAD}%
                           \def\s@ze{8}%
                           \loadsmallfonts\viiipttrue\fi 
                      \else\def\HEAD{fourteen}%
                           \def\smHEAD{ten}%
                           \def\vsHEAD{seven}%
                           \ifxivpt\else\xdef\f@ntsize{\HEAD}%
                           \def\m@g{2}\def\s@ze{14.4}%
                           \loadheadfonts\xivpttrue\fi 
                           \ifxpt\else\xdef\f@ntsize{\smHEAD}%
                           \def\s@ze{10}%
                           \loadxptfonts\xpttrue\fi 
                           \ifviipt\else\xdef\f@ntsize{\vsHEAD}%
                           \def\s@ze{7}%
                           \loadviiptfonts\viipttrue\fi 
                \ifnum#1=14\else 
                \message{Header size should be 20, 17 or 14 point 
                              will now default to 14pt}\fi 
                \fi\fi\headfonts} 
%
%
\def\textsize#1#2{\def\textb@seline{#2}%
                 \ifnum#1=12\def\TEXT{twelve}%
                           \def\smTEXT{eight}%
                           \def\vsTEXT{six}%
                           \ifxiipt\else\xdef\f@ntsize{\TEXT}%
                           \def\m@g{1}\def\s@ze{12}%
                           \loadxiiptfonts\xiipttrue\fi 
                           \ifviiipt\else\xdef\f@ntsize{\smTEXT}%
                           \def\s@ze{8}%
                           \loadsmallfonts\viiipttrue\fi 
                           \ifvipt\else\xdef\f@ntsize{\vsTEXT}%
                           \def\s@ze{6}%
                           \loadviptfonts\vipttrue\fi 
                      \else 
                \ifnum#1=11\def\TEXT{eleven}%
                           \def\smTEXT{seven}%
                           \def\vsTEXT{five}%
                           \ifxipt\else\xdef\f@ntsize{\TEXT}%
                           \def\s@ze{11}%
                           \loadxiptfonts\xipttrue\fi 
                           \ifviipt\else\xdef\f@ntsize{\smTEXT}%
                           \loadviiptfonts\viipttrue\fi 
                           \ifvpt\else\xdef\f@ntsize{\vsTEXT}%
                           \def\s@ze{5}%
                           \loadvptfonts\vpttrue\fi 
                      \else\def\TEXT{ten}%
                           \def\smTEXT{seven}%
                           \def\vsTEXT{five}%
                           \ifxpt\else\xdef\f@ntsize{\TEXT}%
                           \loadxptfonts\xpttrue\fi 
                           \ifviipt\else\xdef\f@ntsize{\smTEXT}%
                           \def\s@ze{7}%
                           \loadviiptfonts\viipttrue\fi 
                           \ifvpt\else\xdef\f@ntsize{\vsTEXT}%
                           \def\s@ze{5}%
                           \loadvptfonts\vpttrue\fi 
                \ifnum#1=10\else 
                \message{Text size should be 12, 11 or 10 point 
                              will now default to 10pt}\fi 
                \fi\fi\textfonts} 
%
%
\def\smallsize#1#2{\def\smallb@seline{#2}%
                 \ifnum#1=10\def\SMALL{ten}%
                           \def\smSMALL{seven}%
                           \def\vsSMALL{five}%
                           \ifxpt\else\xdef\f@ntsize{\SMALL}%
                           \loadxptfonts\xpttrue\fi 
                           \ifviipt\else\xdef\f@ntsize{\smSMALL}%
                           \def\s@ze{7}%
                           \loadviiptfonts\viipttrue\fi 
                           \ifvpt\else\xdef\f@ntsize{\vsSMALL}%
                           \def\s@ze{5}%
                           \loadvptfonts\vpttrue\fi 
                       \else 
                 \ifnum#1=9\def\SMALL{nine}%
                           \def\smSMALL{six}%
                           \def\vsSMALL{five}%
                           \ifixpt\else\xdef\f@ntsize{\SMALL}%
                           \def\s@ze{9}%
                           \loadsmallfonts\ixpttrue\fi 
                           \ifvipt\else\xdef\f@ntsize{\smSMALL}%
                           \def\s@ze{6}%
                           \loadviptfonts\vipttrue\fi 
                           \ifvpt\else\xdef\f@ntsize{\vsSMALL}%
                           \def\s@ze{5}%
                           \loadvptfonts\vpttrue\fi 
                       \else 
                           \def\SMALL{eight}%
                           \def\smSMALL{six}%
                           \def\vsSMALL{five}%
                           \ifviiipt\else\xdef\f@ntsize{\SMALL}%
                           \def\s@ze{8}%
                           \loadsmallfonts\viiipttrue\fi 
                           \ifvipt\else\xdef\f@ntsize{\smSMALL}%
                           \def\s@ze{6}%
                           \loadviptfonts\vipttrue\fi 
                           \ifvpt\else\xdef\f@ntsize{\vsSMALL}%
                           \def\s@ze{5}%
                           \loadvptfonts\vpttrue\fi 
                 \ifnum#1=8\else\message{Small size should be 10, 9 or  
                            8 point will now default to 8pt}\fi 
                \fi\fi\smallfonts} 
\def\F@nt{\expandafter\font\csname} 
\def\Sk@w{\expandafter\skewchar\csname} 
\def\@nd{\endcsname} 
\def\@step#1{ scaled \magstep#1} 
\def\@half{ scaled \magstephalf} 
\def\@t#1{ at #1pt} 
%
%
\def\loadheadfonts{\bigf@nts 
\F@nt \f@ntsize bdi\@nd=cmmib10 \@t{\s@ze}%
\Sk@w \f@ntsize bdi\@nd='177 
\F@nt \f@ntsize bsy\@nd=cmbsy10 \@t{\s@ze}%
\Sk@w \f@ntsize bsy\@nd='60 
\F@nt \f@ntsize bss\@nd=cmssbx10 \@t{\s@ze}} 
%
%
\def\loadxiiptfonts{\bigf@nts 
\F@nt \f@ntsize bdi\@nd=cmmib10 \@step{\m@g}%
\Sk@w \f@ntsize bdi\@nd='177 
\F@nt \f@ntsize bsy\@nd=cmbsy10 \@step{\m@g}%
\Sk@w \f@ntsize bsy\@nd='60 
\F@nt \f@ntsize bss\@nd=cmssbx10 \@step{\m@g}} 
%
%
\def\loadxiptfonts{%
\font\elevenrm=cmr10 \@half 
\font\eleveni=cmmi10 \@half 
\skewchar\eleveni='177 
\font\elevensy=cmsy10 \@half 
\skewchar\elevensy='60 
\font\elevenex=cmex10 \@half 
\font\elevenit=cmti10 \@half 
\font\elevensl=cmsl10 \@half 
\font\elevenbf=cmbx10 \@half 
\font\eleventt=cmtt10 \@half 
\ifams\font\elevenmsa=msam10 \@half 
\font\elevenmsb=msbm10 \@half\else\fi 
\font\elevenbdi=cmmib10 \@half 
\skewchar\elevenbdi='177 
\font\elevenbsy=cmbsy10 \@half 
\skewchar\elevenbsy='60 
\font\elevenbss=cmssbx10 \@half} 
%
%
\def\loadxptfonts{%
\font\tenbdi=cmmib10 
\skewchar\tenbdi='177 
\font\tenbsy=cmbsy10  
\skewchar\tenbsy='60 
\ifams\font\tenmsa=msam10  
\font\tenmsb=msbm10\else\fi 
\font\tenbss=cmssbx10}%
%
%
\def\loadsmallfonts{\smallf@nts 
\ifams 
\F@nt \f@ntsize ex\@nd=cmex\s@ze 
\else 
\F@nt \f@ntsize ex\@nd=cmex10\fi 
\F@nt \f@ntsize it\@nd=cmti\s@ze 
\F@nt \f@ntsize sl\@nd=cmsl\s@ze 
\F@nt \f@ntsize tt\@nd=cmtt\s@ze} 
%
%
\def\loadviiptfonts{%
\font\sevenit=cmti7 
\font\sevensl=cmsl8 at 7pt 
\ifams\font\sevenmsa=msam7  
\font\sevenmsb=msbm7 
\font\sevenex=cmex7 
\font\sevenbsy=cmbsy7 
\font\sevenbdi=cmmib7\else 
\font\sevenex=cmex10 
\font\sevenbsy=cmbsy10 at 7pt 
\font\sevenbdi=cmmib10 at 7pt\fi 
\skewchar\sevenbsy='60 
\skewchar\sevenbdi='177 
\font\sevenbss=cmssbx10 at 7pt}%
%
%
\def\loadviptfonts{\smallf@nts 
\ifams\font\sixex=cmex7 at 6pt\else 
\font\sixex=cmex10\fi 
\font\sixit=cmti7 at 6pt} 
%
%
\def\loadvptfonts{%
\font\fiveit=cmti7 at 5pt 
\ifams\font\fiveex=cmex7 at 5pt 
\font\fivebdi=cmmib5 
\font\fivebsy=cmbsy5 
\font\fivemsa=msam5  
\font\fivemsb=msbm5\else 
\font\fiveex=cmex10 
\font\fivebdi=cmmib10 at 5pt 
\font\fivebsy=cmbsy10 at 5pt\fi 
\skewchar\fivebdi='177 
\skewchar\fivebsy='60 
\font\fivebss=cmssbx10 at 5pt} 
\def\bigf@nts{%
\F@nt \f@ntsize rm\@nd=cmr10 \@step{\m@g}%
\F@nt \f@ntsize i\@nd=cmmi10 \@step{\m@g}%
\Sk@w \f@ntsize i\@nd='177 
\F@nt \f@ntsize sy\@nd=cmsy10 \@step{\m@g}%
\Sk@w \f@ntsize sy\@nd='60 
\F@nt \f@ntsize ex\@nd=cmex10 \@step{\m@g}%
\F@nt \f@ntsize it\@nd=cmti10 \@step{\m@g}%
\F@nt \f@ntsize sl\@nd=cmsl10 \@step{\m@g}%
\F@nt \f@ntsize bf\@nd=cmbx10 \@step{\m@g}%
\F@nt \f@ntsize tt\@nd=cmtt10 \@step{\m@g}%
\ifams 
\F@nt \f@ntsize msa\@nd=msam10 \@step{\m@g}%
\F@nt \f@ntsize msb\@nd=msbm10 \@step{\m@g}\else\fi} 
\def\smallf@nts{%
\F@nt \f@ntsize rm\@nd=cmr\s@ze 
\F@nt \f@ntsize i\@nd=cmmi\s@ze  
\Sk@w \f@ntsize i\@nd='177 
\F@nt \f@ntsize sy\@nd=cmsy\s@ze 
\Sk@w \f@ntsize sy\@nd='60 
\F@nt \f@ntsize bf\@nd=cmbx\s@ze  
\ifams 
\F@nt \f@ntsize bdi\@nd=cmmib\s@ze  
\F@nt \f@ntsize bsy\@nd=cmbsy\s@ze  
\F@nt \f@ntsize msa\@nd=msam\s@ze  
\F@nt \f@ntsize msb\@nd=msbm\s@ze 
\else 
\F@nt \f@ntsize bdi\@nd=cmmib10 \@t{\s@ze}%
\F@nt \f@ntsize bsy\@nd=cmbsy10 \@t{\s@ze}\fi  
\Sk@w \f@ntsize bdi\@nd='177 
\Sk@w \f@ntsize bsy\@nd='60 
\F@nt \f@ntsize bss\@nd=cmssbx10 \@t{\s@ze}}%
%
%
\def\headfonts{%
\textfont0=\csname\HEAD rm\@nd         
\scriptfont0=\csname\smHEAD rm\@nd 
\scriptscriptfont0=\csname\vsHEAD rm\@nd 
\def\rm{\fam0\csname\HEAD rm\@nd 
\def\sc{\csname\smHEAD rm\@nd}}%
\textfont1=\csname\HEAD i\@nd          
\scriptfont1=\csname\smHEAD i\@nd 
\scriptscriptfont1=\csname\vsHEAD i\@nd 
\textfont2=\csname\HEAD sy\@nd         
\scriptfont2=\csname\smHEAD sy\@nd 
\scriptscriptfont2=\csname\vsHEAD sy\@nd 
\textfont3=\csname\HEAD ex\@nd         
\scriptfont3=\csname\smHEAD ex\@nd 
\scriptscriptfont3=\csname\smHEAD ex\@nd 
\textfont\itfam=\csname\HEAD it\@nd    
\scriptfont\itfam=\csname\smHEAD it\@nd 
\scriptscriptfont\itfam=\csname\vsHEAD it\@nd 
\def\it{\fam\itfam\csname\HEAD it\@nd 
\def\sc{\csname\smHEAD it\@nd}}%
\textfont\slfam=\csname\HEAD sl\@nd    
\def\sl{\fam\slfam\csname\HEAD sl\@nd 
\def\sc{\csname\smHEAD sl\@nd}}%
\textfont\bffam=\csname\HEAD bf\@nd    
\scriptfont\bffam=\csname\smHEAD bf\@nd 
\scriptscriptfont\bffam=\csname\vsHEAD bf\@nd 
\def\bf{\fam\bffam\csname\HEAD bf\@nd 
\def\sc{\csname\smHEAD bf\@nd}}%
\textfont\ttfam=\csname\HEAD tt\@nd    
\def\tt{\fam\ttfam\csname\HEAD tt\@nd}%
\textfont\bdifam=\csname\HEAD bdi\@nd  
\scriptfont\bdifam=\csname\smHEAD bdi\@nd 
\scriptscriptfont\bdifam=\csname\vsHEAD bdi\@nd 
\def\bdi{\fam\bdifam\csname\HEAD bdi\@nd}%
\textfont\bsyfam=\csname\HEAD bsy\@nd  
\scriptfont\bsyfam=\csname\smHEAD bsy\@nd 
\def\bsy{\fam\bsyfam\csname\HEAD bsy\@nd}%
\textfont\bssfam=\csname\HEAD bss\@nd  
\scriptfont\bssfam=\csname\smHEAD bss\@nd 
\scriptscriptfont\bssfam=\csname\vsHEAD bss\@nd 
\def\bss{\fam\bssfam\csname\HEAD bss\@nd}%
\ifams 
\textfont\msafam=\csname\HEAD msa\@nd  
\scriptfont\msafam=\csname\smHEAD msa\@nd 
\scriptscriptfont\msafam=\csname\vsHEAD msa\@nd 
\textfont\msbfam=\csname\HEAD msb\@nd  
\scriptfont\msbfam=\csname\smHEAD msb\@nd 
\scriptscriptfont\msbfam=\csname\vsHEAD msb\@nd 
\else\fi 
\normalbaselineskip=\headb@seline pt%
\setbox\strutbox=\hbox{\vrule height.7\normalbaselineskip  
depth.3\baselineskip width0pt}%
\def\sc{\csname\smHEAD rm\@nd}\normalbaselines\bf} 
%
%
\def\textfonts{%
\textfont0=\csname\TEXT rm\@nd         
\scriptfont0=\csname\smTEXT rm\@nd 
\scriptscriptfont0=\csname\vsTEXT rm\@nd 
\def\rm{\fam0\csname\TEXT rm\@nd 
\def\sc{\csname\smTEXT rm\@nd}}%
\textfont1=\csname\TEXT i\@nd          
\scriptfont1=\csname\smTEXT i\@nd 
\scriptscriptfont1=\csname\vsTEXT i\@nd 
\textfont2=\csname\TEXT sy\@nd         
\scriptfont2=\csname\smTEXT sy\@nd 
\scriptscriptfont2=\csname\vsTEXT sy\@nd 
\textfont3=\csname\TEXT ex\@nd         
\scriptfont3=\csname\smTEXT ex\@nd 
\scriptscriptfont3=\csname\smTEXT ex\@nd 
\textfont\itfam=\csname\TEXT it\@nd    
\scriptfont\itfam=\csname\smTEXT it\@nd 
\scriptscriptfont\itfam=\csname\vsTEXT it\@nd 
\def\it{\fam\itfam\csname\TEXT it\@nd 
\def\sc{\csname\smTEXT it\@nd}}%
\textfont\slfam=\csname\TEXT sl\@nd    
\def\sl{\fam\slfam\csname\TEXT sl\@nd 
\def\sc{\csname\smTEXT sl\@nd}}%
\textfont\bffam=\csname\TEXT bf\@nd    
\scriptfont\bffam=\csname\smTEXT bf\@nd 
\scriptscriptfont\bffam=\csname\vsTEXT bf\@nd 
\def\bf{\fam\bffam\csname\TEXT bf\@nd 
\def\sc{\csname\smTEXT bf\@nd}}%
\textfont\ttfam=\csname\TEXT tt\@nd    
\def\tt{\fam\ttfam\csname\TEXT tt\@nd}%
\textfont\bdifam=\csname\TEXT bdi\@nd  
\scriptfont\bdifam=\csname\smTEXT bdi\@nd 
\scriptscriptfont\bdifam=\csname\vsTEXT bdi\@nd 
\def\bdi{\fam\bdifam\csname\TEXT bdi\@nd}%
\textfont\bsyfam=\csname\TEXT bsy\@nd  
\scriptfont\bsyfam=\csname\smTEXT bsy\@nd 
\def\bsy{\fam\bsyfam\csname\TEXT bsy\@nd}%
\textfont\bssfam=\csname\TEXT bss\@nd  
\scriptfont\bssfam=\csname\smTEXT bss\@nd 
\scriptscriptfont\bssfam=\csname\vsTEXT bss\@nd 
\def\bss{\fam\bssfam\csname\TEXT bss\@nd}%
\ifams 
\textfont\msafam=\csname\TEXT msa\@nd  
\scriptfont\msafam=\csname\smTEXT msa\@nd 
\scriptscriptfont\msafam=\csname\vsTEXT msa\@nd 
\textfont\msbfam=\csname\TEXT msb\@nd  
\scriptfont\msbfam=\csname\smTEXT msb\@nd 
\scriptscriptfont\msbfam=\csname\vsTEXT msb\@nd 
\else\fi 
\normalbaselineskip=\textb@seline pt 
\setbox\strutbox=\hbox{\vrule height.7\normalbaselineskip  
depth.3\baselineskip width0pt}%
\everymath{}%
\def\sc{\csname\smTEXT rm\@nd}\normalbaselines\rm} 
%
%
\def\smallfonts{%
\textfont0=\csname\SMALL rm\@nd         
\scriptfont0=\csname\smSMALL rm\@nd 
\scriptscriptfont0=\csname\vsSMALL rm\@nd 
\def\rm{\fam0\csname\SMALL rm\@nd 
\def\sc{\csname\smSMALL rm\@nd}}%
\textfont1=\csname\SMALL i\@nd          
\scriptfont1=\csname\smSMALL i\@nd 
\scriptscriptfont1=\csname\vsSMALL i\@nd 
\textfont2=\csname\SMALL sy\@nd         
\scriptfont2=\csname\smSMALL sy\@nd 
\scriptscriptfont2=\csname\vsSMALL sy\@nd 
\textfont3=\csname\SMALL ex\@nd         
\scriptfont3=\csname\smSMALL ex\@nd 
\scriptscriptfont3=\csname\smSMALL ex\@nd 
\textfont\itfam=\csname\SMALL it\@nd    
\scriptfont\itfam=\csname\smSMALL it\@nd 
\scriptscriptfont\itfam=\csname\vsSMALL it\@nd 
\def\it{\fam\itfam\csname\SMALL it\@nd 
\def\sc{\csname\smSMALL it\@nd}}%
\textfont\slfam=\csname\SMALL sl\@nd    
\def\sl{\fam\slfam\csname\SMALL sl\@nd 
\def\sc{\csname\smSMALL sl\@nd}}%
\textfont\bffam=\csname\SMALL bf\@nd    
\scriptfont\bffam=\csname\smSMALL bf\@nd 
\scriptscriptfont\bffam=\csname\vsSMALL bf\@nd 
\def\bf{\fam\bffam\csname\SMALL bf\@nd 
\def\sc{\csname\smSMALL bf\@nd}}%
\textfont\ttfam=\csname\SMALL tt\@nd    
\def\tt{\fam\ttfam\csname\SMALL tt\@nd}%
\textfont\bdifam=\csname\SMALL bdi\@nd  
\scriptfont\bdifam=\csname\smSMALL bdi\@nd 
\scriptscriptfont\bdifam=\csname\vsSMALL bdi\@nd 
\def\bdi{\fam\bdifam\csname\SMALL bdi\@nd}%
\textfont\bsyfam=\csname\SMALL bsy\@nd  
\scriptfont\bsyfam=\csname\smSMALL bsy\@nd 
\def\bsy{\fam\bsyfam\csname\SMALL bsy\@nd}%
\textfont\bssfam=\csname\SMALL bss\@nd  
\scriptfont\bssfam=\csname\smSMALL bss\@nd 
\scriptscriptfont\bssfam=\csname\vsSMALL bss\@nd 
\def\bss{\fam\bssfam\csname\SMALL bss\@nd}%
\ifams 
\textfont\msafam=\csname\SMALL msa\@nd  
\scriptfont\msafam=\csname\smSMALL msa\@nd 
\scriptscriptfont\msafam=\csname\vsSMALL msa\@nd 
\textfont\msbfam=\csname\SMALL msb\@nd  
\scriptfont\msbfam=\csname\smSMALL msb\@nd 
\scriptscriptfont\msbfam=\csname\vsSMALL msb\@nd 
\else\fi 
\normalbaselineskip=\smallb@seline pt%
\setbox\strutbox=\hbox{\vrule height.7\normalbaselineskip  
depth.3\baselineskip width0pt}%
\everymath{}%
\def\sc{\csname\smSMALL rm\@nd}\normalbaselines\rm}%
\everydisplay{\indenteddisplay 
   \gdef\labeltype{\eqlabel}}%
%
%
\def\hexnumber@#1{\ifcase#1 0\or 1\or 2\or 3\or 4\or 5\or 6\or 7\or 8\or 
 9\or A\or B\or C\or D\or E\or F\fi} 
\edef\bffam@{\hexnumber@\bffam} 
\edef\bdifam@{\hexnumber@\bdifam} 
\edef\bsyfam@{\hexnumber@\bsyfam} 
\def\undefine#1{\let#1\undefined} 
\def\newsymbol#1#2#3#4#5{\let\next@\relax 
 \ifnum#2=\thr@@\let\next@\bdifam@\else 
 \ifams 
 \ifnum#2=\@ne\let\next@\msafam@\else 
 \ifnum#2=\tw@\let\next@\msbfam@\fi\fi 
 \fi\fi 
 \mathchardef#1="#3\next@#4#5} 
\def\mathhexbox@#1#2#3{\relax 
 \ifmmode\mathpalette{}{\m@th\mathchar"#1#2#3}%
 \else\leavevmode\hbox{$\m@th\mathchar"#1#2#3$}\fi} 

\def\bi#1{{\fam\bdifam\relax#1}} 
%
%
\ifams\input amsmacro\fi 
%
%
\newsymbol\bitGamma 3000 
\newsymbol\bitDelta 3001 
\newsymbol\bitTheta 3002 
\newsymbol\bitLambda 3003 
\newsymbol\bitXi 3004 
\newsymbol\bitPi 3005 
\newsymbol\bitSigma 3006 
\newsymbol\bitUpsilon 3007 
\newsymbol\bitPhi 3008 
\newsymbol\bitPsi 3009 
\newsymbol\bitOmega 300A 
\newsymbol\balpha 300B 
\newsymbol\bbeta 300C 
\newsymbol\bgamma 300D 
\newsymbol\bdelta 300E 
\newsymbol\bepsilon 300F 
\newsymbol\bzeta 3010 
\newsymbol\bfeta 3011 
\newsymbol\btheta 3012 
\newsymbol\biota 3013 
\newsymbol\bkappa 3014 
\newsymbol\blambda 3015 
\newsymbol\bmu 3016 
\newsymbol\bnu 3017 
\newsymbol\bxi 3018 
\newsymbol\bpi 3019 
\newsymbol\brho 301A 
\newsymbol\bsigma 301B 
\newsymbol\btau 301C 
\newsymbol\bupsilon 301D 
\newsymbol\bphi 301E 
\newsymbol\bchi 301F 
\newsymbol\bpsi 3020 
\newsymbol\bomega 3021 
\newsymbol\bvarepsilon 3022 
\newsymbol\bvartheta 3023 
\newsymbol\bvaromega 3024 
\newsymbol\bvarrho 3025 
\newsymbol\bvarzeta 3026 
\newsymbol\bvarphi 3027 
\newsymbol\bpartial 3040 
\newsymbol\bell 3060 
\newsymbol\bimath 307B 
\newsymbol\bjmath 307C 
\mathchardef\binfty "0\bsyfam@31 
\mathchardef\bnabla "0\bsyfam@72 
\mathchardef\bdot "2\bsyfam@01 
\mathchardef\bGamma "0\bffam@00 
\mathchardef\bDelta "0\bffam@01 
\mathchardef\bTheta "0\bffam@02 
\mathchardef\bLambda "0\bffam@03 
\mathchardef\bXi "0\bffam@04 
\mathchardef\bPi "0\bffam@05 
\mathchardef\bSigma "0\bffam@06 
\mathchardef\bUpsilon "0\bffam@07 
\mathchardef\bPhi "0\bffam@08 
\mathchardef\bPsi "0\bffam@09 
\mathchardef\bOmega "0\bffam@0A 
\mathchardef\itGamma "0100 
\mathchardef\itDelta "0101 
\mathchardef\itTheta "0102 
\mathchardef\itLambda "0103 
\mathchardef\itXi "0104 
\mathchardef\itPi "0105 
\mathchardef\itSigma "0106 
\mathchardef\itUpsilon "0107 
\mathchardef\itPhi "0108 
\mathchardef\itPsi "0109 
\mathchardef\itOmega "010A 
\mathchardef\Gamma "0000 
\mathchardef\Delta "0001 
\mathchardef\Theta "0002 
\mathchardef\Lambda "0003 
\mathchardef\Xi "0004 
\mathchardef\Pi "0005 
\mathchardef\Sigma "0006 
\mathchardef\Upsilon "0007 
\mathchardef\Phi "0008 
\mathchardef\Psi "0009 
\mathchardef\Omega "000A 
%
%
\newcount\firstpage  \firstpage=1  
\newcount\jnl                      
\newcount\secno                    
\newcount\subno                    
\newcount\subsubno                 
\newcount\appno                    
\newcount\tabno                    
\newcount\figno                    
\newcount\countno                  
\newcount\refno                    
\newcount\eqlett     \eqlett=97    
\newif\ifletter 
\newif\ifwide 
\newif\ifnotfull 
\newif\ifaligned 
\newif\ifnumbysec   
\newif\ifappendix 
\newif\ifnumapp 
\newif\ifssf 
\newif\ifppt 
\newdimen\t@bwidth 
\newdimen\c@pwidth 
\newdimen\digitwidth                    
\newdimen\argwidth                      
\newdimen\secindent    \secindent=5pc   
\newdimen\textind    \textind=16pt      
\newdimen\tempval                       
\newskip\beforesecskip 
\def\beforesecspace{\vskip\beforesecskip\relax} 
\newskip\beforesubskip 
\def\beforesubspace{\vskip\beforesubskip\relax} 
\newskip\beforesubsubskip 
\def\beforesubsubspace{\vskip\beforesubsubskip\relax} 
\newskip\secskip 
\def\secspace{\vskip\secskip\relax} 
\newskip\subskip 
\def\subspace{\vskip\subskip\relax} 
\newskip\insertskip 
\def\insertspace{\vskip\insertskip\relax} 
\def\sp@ce{\ifx\next*\let\next=\@ssf 
               \else\let\next=\@nossf\fi\next} 
\def\@ssf#1{\nobreak\secspace\global\ssftrue\nobreak} 
\def\@nossf{\nobreak\secspace\nobreak\noindent\ignorespaces} 
\def\subsp@ce{\ifx\next*\let\next=\@sssf 
               \else\let\next=\@nosssf\fi\next} 
\def\@sssf#1{\nobreak\subspace\global\ssftrue\nobreak} 
\def\@nosssf{\nobreak\subspace\nobreak\noindent\ignorespaces} 
\beforesecskip=24pt plus12pt minus8pt 
\beforesubskip=12pt plus6pt minus4pt 
\beforesubsubskip=12pt plus6pt minus4pt 
\secskip=12pt plus 2pt minus 2pt 
\subskip=6pt plus3pt minus2pt 
\insertskip=18pt plus6pt minus6pt%
\fontdimen16\tensy=2.7pt 
\fontdimen17\tensy=2.7pt 
%
%
\def\eqlabel{(\ifappendix\applett 
               \ifnumbysec\ifnum\secno>0 \the\secno\fi.\fi 
               \else\ifnumbysec\the\secno.\fi\fi\the\countno)} 
\def\seclabel{\ifappendix\ifnumapp\else\applett\fi 
    \ifnum\secno>0 \the\secno 
    \ifnumbysec\ifnum\subno>0.\the\subno\fi\fi\fi 
    \else\the\secno\fi\ifnum\subno>0.\the\subno 
         \ifnum\subsubno>0.\the\subsubno\fi\fi} 
\def\tablabel{\ifappendix\applett\fi\the\tabno} 
\def\figlabel{\ifappendix\applett\fi\the\figno} 
\def\gac{\global\advance\countno by 1} 
%
%
 
\def\vfootnote#1{\insert\footins\bgroup 
\interlinepenalty=\interfootnotelinepenalty 
\splittopskip=\ht\strutbox 
\splitmaxdepth=\dp\strutbox \floatingpenalty=20000 
\leftskip=0pt \rightskip=0pt \spaceskip=0pt \xspaceskip=0pt%
\noindent\smallfonts\rm #1\ \ignorespaces\footstrut\futurelet\next\fo@t} 
%
%
\def\endinsert{\egroup 
    \if@mid \dimen@=\ht0 \advance\dimen@ by\dp0 
       \advance\dimen@ by12\p@ \advance\dimen@ by\pagetotal 
       \ifdim\dimen@>\pagegoal \@midfalse\p@gefalse\fi\fi 
    \if@mid \insertspace \box0 \par \ifdim\lastskip<\insertskip 
    \removelastskip \penalty-200 \insertspace \fi 
    \else\insert\topins{\penalty100 
       \splittopskip=0pt \splitmaxdepth=\maxdimen  
       \floatingpenalty=0 
       \ifp@ge \dimen@=\dp0 
       \vbox to\vsize{\unvbox0 \kern-\dimen@}%
       \else\box0\nobreak\insertspace\fi}\fi\endgroup}    
%
%
%
\def\ind{\hbox to \secindent{\hfill}} 
%
%

%
%
 
%
%
\def\indeqn#1{\alignedfalse\displ@y\halign{\hbox to \displaywidth 
    {$\ind\@lign\displaystyle##\hfil$}\crcr #1\crcr}} 
%
%
\def\indalign#1{\alignedtrue\displ@y \tabskip=0pt  
  \halign to\displaywidth{\ind$\@lign\displaystyle{##}$\tabskip=0pt 
    &$\@lign\displaystyle{{}##}$\hfill\tabskip=\centering 
    &\llap{$\@lign\hbox{\rm##}$}\tabskip=0pt\crcr 
    #1\crcr}} 
\def\fl{{\hskip-\secindent}} 
\def\indenteddisplay#1$${\indispl@y{#1 }} 
\def\indispl@y#1{\disptest#1\eqalignno\eqalignno\disptest} 
\def\disptest#1\eqalignno#2\eqalignno#3\disptest{%
    \ifx#3\eqalignno 
    \indalign#2%
    \else\indeqn{#1}\fi$$} 
%
%
 
%
%
 
%
%
 
%
%
 
%
%

\def\ns{\noalign{\vskip-3pt}}

%
 
%
%
\def\bhbar{\rlap{\kern1pt\raise.4ex\hbox{\bf\char'40}}\bi{h}} 

\def\d{{\rm d}}

\def\frac#1#2{{#1\over#2}} 
\ifams 
\def\lap{\lesssim} 
\def\gap{\gtrsim}

\let\leq=\leqslant

\else

\def\gap{\;\lower3pt\hbox{$\buildrel > \over \sim$}\;}%
\def\lap{\;\lower3pt\hbox{$\buildrel < \over \sim$}\;}\fi 
 
\chardef\ii="10 
\def\tqs{\hbox to 25pt{\hfil}}

\def\Bbbone{1\kern-.22em {\rm l}} 
%
%
\def\rp{\raise8pt\hbox{$\scriptstyle\prime$}} 
%
%
%
%

%
%
\def\[#1\]{\setbox0=\hbox{$\dsty#1$}\argwidth=\wd0 
    \setbox0=\hbox{$\left[\box0\right]$}\advance\argwidth by -\wd0 
    \left[\kern.3\argwidth\box0\kern.3\argwidth\right]} 
%
%
\def\lsb#1\rsb{\setbox0=\hbox{$#1$}\argwidth=\wd0 
    \setbox0=\hbox{$\left[\box0\right]$}\advance\argwidth by -\wd0 
    \left[\kern.3\argwidth\box0\kern.3\argwidth\right]} 
%
 
%
%
 
%
\def\pt(#1){({\it #1\/})} 
\let\dsty=\displaystyle

%
%
\def\reactions#1{\vskip 12pt plus2pt minus2pt%
\vbox{\hbox{\kern\secindent\vrule\kern12pt%
\vbox{\kern0.5pt\vbox{\hsize=24pc\parindent=0pt\smallfonts\rm NUCLEAR  
REACTIONS\strut\quad #1\strut}\kern0.5pt}\kern12pt\vrule}}} 
%
%
\def\slashchar#1{\setbox0=\hbox{$#1$}\dimen0=\wd0%
\setbox1=\hbox{/}\dimen1=\wd1%
\ifdim\dimen0>\dimen1%
\rlap{\hbox to \dimen0{\hfil/\hfil}}#1\else                                         
\rlap{\hbox to \dimen1{\hfil$#1$\hfil}}/\fi} 
%
%
\def\textindent#1{\noindent\hbox to \parindent{#1\hss}\ignorespaces} 
%
%
\def\opencirc{\raise1pt\hbox{$\scriptstyle{\bigcirc}$}} 
 
\ifams 
\def\opensqr{\hbox{$\square$}} 
 
\def\opentridown{\hbox{$\triangledown$}}

\else 
\def\opensqr{\vbox{\hrule height.4pt\hbox{\vrule width.4pt height3.5pt 
    \kern3.5pt\vrule width.4pt}\hrule height.4pt}} 
 
\def\opentridown{\raise1pt\hbox{$\scriptstyle\bigtriangledown$}}

\fi

%
%
\def\m@th{\mathsurround=0pt} 
%
%
\def\cases#1{%
\left\{\,\vcenter{\normalbaselines\openup1\jot\m@th%
     \ialign{$\displaystyle##\hfil$&\rm\tqs##\hfil\crcr#1\crcr}}\right.}%
%
%
\def\oldcases#1{\left\{\,\vcenter{\normalbaselines\m@th 
    \ialign{$##\hfil$&\rm\quad##\hfil\crcr#1\crcr}}\right.} 
%
%
\def\numcases#1{\left\{\,\vcenter{\baselineskip=15pt\m@th%
     \ialign{$\displaystyle##\hfil$&\rm\tqs##\hfil 
     \crcr#1\crcr}}\right.\hfill 
     \vcenter{\baselineskip=15pt\m@th%
     \ialign{\rlap{$\phantom{\displaystyle##\hfil}$}\tabskip=0pt&\en 
     \rlap{\phantom{##\hfil}}\crcr#1\crcr}}} 
\def\ptnumcases#1{\left\{\,\vcenter{\baselineskip=15pt\m@th%
     \ialign{$\displaystyle##\hfil$&\rm\tqs##\hfil 
     \crcr#1\crcr}}\right.\hfill 
     \vcenter{\baselineskip=15pt\m@th%
     \ialign{\rlap{$\phantom{\displaystyle##\hfil}$}\tabskip=0pt&\enpt 
     \rlap{\phantom{##\hfil}}\crcr#1\crcr}}\global\eqlett=97 
     \global\advance\countno by 1} 
%
%
\def\eq(#1){\ifaligned\@mp(#1)\else\hfill\llap{{\rm (#1)}}\fi} 
\def\ceq(#1){\ns\ns\ifaligned\@mp\fi\eq(#1)\cr\ns\ns} 
\def\eqpt(#1#2){\ifaligned\@mp(#1{\it #2\/}) 
                    \else\hfill\llap{{\rm (#1{\it #2\/})}}\fi} 
\let\eqno=\eq 
%
%
\countno=1 
 
\def\aleq{&\rm(\ifappendix\applett 
               \ifnumbysec\ifnum\secno>0 \the\secno\fi.\fi 
               \else\ifnumbysec\the\secno.\fi\fi\the\countno} 
\def\noaleq{\hfill\llap\bgroup\rm(\ifappendix\applett 
               \ifnumbysec\ifnum\secno>0 \the\secno\fi.\fi 
               \else\ifnumbysec\the\secno.\fi\fi\the\countno} 
\def\@mp{&} 
\def\en{\ifaligned\aleq)\else\noaleq)\egroup\fi\gac} 
\def\cen{\ns\ns\ifaligned\@mp\fi\en\cr\ns\ns} 
\def\enpt{\ifaligned\aleq{\it\char\the\eqlett})\else 
    \noaleq{\it\char\the\eqlett})\egroup\fi 
    \global\advance\eqlett by 1} 
\def\endpt{\ifaligned\aleq{\it\char\the\eqlett})\else 
    \noaleq{\it\char\the\eqlett})\egroup\fi 
    \global\eqlett=97\gac} 
%
%

\def\JPA{{\it J. Phys. A: Math. Gen.}} 

 

%
%

\def\JP{{\it J. Physique\/}}

\def\PR{{\it Phys. Rev.}}

\def\ZP{{\it Z. Phys.}} 
\headline={\ifodd\pageno{\ifnum\pageno=\firstpage\hfill 
   \else\rrhead\fi}\else\lrhead\fi} 
\def\rrhead{\textfonts\hskip\secindent\it 
    \shorttitle\hfill\rm\folio} 
\def\lrhead{\textfonts\hbox to\secindent{\rm\folio\hss}%
    \it\aunames\hss} 
\footline={\ifnum\pageno=\firstpage \hfill\textfonts\rm\folio\fi} 
\def\@rticle#1#2{\vglue.5pc 
    {\parindent=\secindent \bf #1\par} 
     \vskip2.5pc 
    {\exhyphenpenalty=10000\hyphenpenalty=10000 
     \baselineskip=18pt\raggedright\noindent 
     \headfonts\bf#2\par}\futurelet\next\sh@rttitle}%
\def\title#1{\gdef\shorttitle{#1} 
    \vglue4pc{\exhyphenpenalty=10000\hyphenpenalty=10000  
    \baselineskip=18pt  
    \raggedright\parindent=0pt 
    \headfonts\bf#1\par}\futurelet\next\sh@rttitle}  

\def\article#1#2{\gdef\shorttitle{#2}\@rticle{#1}{#2}}  
\def\review#1{\gdef\shorttitle{#1}%
    \@rticle{REVIEW \ifpbm\else ARTICLE\fi}{#1}} 
\def\topical#1{\gdef\shorttitle{#1}%
    \@rticle{TOPICAL REVIEW}{#1}} 
\def\comment#1{\gdef\shorttitle{#1}%
    \@rticle{COMMENT}{#1}} 
\def\note#1{\gdef\shorttitle{#1}%
    \@rticle{NOTE}{#1}} 
\def\prelim#1{\gdef\shorttitle{#1}%
    \@rticle{PRELIMINARY COMMUNICATION}{#1}} 
\def\letter#1{\gdef\shorttitle{Letter to the Editor}%
     \gdef\aunames{Letter to the Editor} 
     \global\lettertrue\ifnum\jnl=7\global\letterfalse\fi 
     \@rticle{LETTER TO THE EDITOR}{#1}} 
\def\sh@rttitle{\ifx\next[\let\next=\sh@rt 
                \else\let\next=\f@ll\fi\next} 
\def\sh@rt[#1]{\gdef\shorttitle{#1}} 
\def\f@ll{} 
\def\author#1{\ifletter\else\gdef\aunames{#1}\fi\vskip1.5pc 
    {\parindent=\secindent   
     \hang\textfonts   
     \ifppt\bf\else\rm\fi#1\par}   
     \ifppt\bigskip\else\smallskip\fi 
     \futurelet\next\@unames} 
\def\@unames{\ifx\next[\let\next=\short@uthor 
                 \else\let\next=\@uthor\fi\next} 
\def\short@uthor[#1]{\gdef\aunames{#1}} 
\def\@uthor{} 
\def\address#1{{\parindent=\secindent 
    \exhyphenpenalty=10000\hyphenpenalty=10000 
\ifppt\textfonts\else\smallfonts\fi\hang\raggedright\rm#1\par}%
    \ifppt\bigskip\fi} 
\def\jl#1{\global\jnl=#1} 
\jl{0}%
\def\journal{\ifnum\jnl=1 J. Phys.\ A: Math.\ Gen.\  
        \else\ifnum\jnl=2 J. Phys.\ B: At.\ Mol.\ Opt.\ Phys.\  
        \else\ifnum\jnl=3 J. Phys.:\ Condens. Matter\  
        \else\ifnum\jnl=4 J. Phys.\ G: Nucl.\ Part.\ Phys.\  
        \else\ifnum\jnl=5 Inverse Problems\  
        \else\ifnum\jnl=6 Class. Quantum Grav.\  
        \else\ifnum\jnl=7 Network\  
        \else\ifnum\jnl=8 Nonlinearity\ 
        \else\ifnum\jnl=9 Quantum Opt.\ 
        \else\ifnum\jnl=10 Waves in Random Media\ 
        \else\ifnum\jnl=11 Pure Appl. Opt.\  
        \else\ifnum\jnl=12 Phys. Med. Biol.\ 
        \else\ifnum\jnl=13 Modelling Simulation Mater.\ Sci.\ Eng.\  
        \else\ifnum\jnl=14 Plasma Phys. Control. Fusion\  
        \else\ifnum\jnl=15 Physiol. Meas.\  
        \else\ifnum\jnl=16 Sov.\ Lightwave Commun.\ 
        \else\ifnum\jnl=17 J. Phys.\ D: Appl.\ Phys.\ 
        \else\ifnum\jnl=18 Supercond.\ Sci.\ Technol.\ 
        \else\ifnum\jnl=19 Semicond.\ Sci.\ Technol.\ 
        \else\ifnum\jnl=20 Nanotechnology\ 
        \else\ifnum\jnl=21 Meas.\ Sci.\ Technol.\  
        \else\ifnum\jnl=22 Plasma Sources Sci.\ Technol.\  
        \else\ifnum\jnl=23 Smart Mater.\ Struct.\  
        \else\ifnum\jnl=24 J.\ Micromech.\ Microeng.\ 
   \else Institute of Physics Publishing\  
   \fi\fi\fi\fi\fi\fi\fi\fi\fi\fi\fi\fi\fi\fi\fi 
   \fi\fi\fi\fi\fi\fi\fi\fi\fi} 
\let\abs=\beginabstract 

\let\endabs=\endabstract 
\def\submitted{\ifppt\noindent\textfonts\rm Submitted to \journal\par 
     \bigskip\fi} 
\def\today{\number\day\ \ifcase\month\or 
     January\or February\or March\or April\or May\or June\or 
     July\or August\or September\or October\or November\or 
     December\fi\space \number\year} 
\def\date{\ifppt\noindent\textfonts\rm  
     Date: \today\par\goodbreak\bigskip\fi} 
%
%
\def\pacs#1{\ifppt\noindent\textfonts\rm  
     PACS number(s): #1\par\bigskip\fi} 
%
 
%
%
\def\section#1{\ifppt\ifnum\secno=0\eject\fi\fi 
    \subno=0\subsubno=0\global\advance\secno by 1 
    \gdef\labeltype{\seclabel}\ifnumbysec\countno=1\fi 
    \goodbreak\beforesecspace\nobreak 
    \noindent{\bf \the\secno. #1}\par\futurelet\next\sp@ce} 
\def\subsection#1{\subsubno=0\global\advance\subno by 1 
     \gdef\labeltype{\seclabel}%
     \ifssf\else\goodbreak\beforesubspace\fi 
     \global\ssffalse\nobreak 
     \noindent{\it \the\secno.\the\subno. #1\par}%
     \futurelet\next\subsp@ce} 
\def\subsubsection#1{\global\advance\subsubno by 1 
     \gdef\labeltype{\seclabel}%
     \ifssf\else\goodbreak\beforesubsubspace\fi 
     \global\ssffalse\nobreak 
     \noindent{\it \the\secno.\the\subno.\the\subsubno. #1}\null.  
     \ignorespaces} 
%
 
%
%
\def\numappendix#1{\ifappendix\ifnumbysec\countno=1\fi\else 
    \countno=1\figno=0\tabno=0\fi 
    \subno=0\global\advance\appno by 1 
    \secno=\appno\gdef\applett{A}\gdef\labeltype{\seclabel}%
    \global\appendixtrue\global\numapptrue 
    \goodbreak\beforesecspace\nobreak 
    \noindent{\bf Appendix \the\appno. #1\par}%
    \futurelet\next\sp@ce} 
\def\numsubappendix#1{\global\advance\subno by 1\subsubno=0 
    \gdef\labeltype{\seclabel}%
    \ifssf\else\goodbreak\beforesubspace\fi 
    \global\ssffalse\nobreak 
    \noindent{\it A\the\appno.\the\subno. #1\par}%
    \futurelet\next\subsp@ce} 
\def\@ppendix#1#2#3{\countno=1\subno=0\subsubno=0\secno=0\figno=0\tabno=0 
    \gdef\applett{#1}\gdef\labeltype{\seclabel}\global\appendixtrue 
    \goodbreak\beforesecspace\nobreak 
    \noindent{\bf Appendix#2#3\par}\futurelet\next\sp@ce} 
\def\Appendix#1{\@ppendix{A}{. }{#1}} 
\def\appendix#1#2{\@ppendix{#1}{ #1. }{#2}} 
\def\App#1{\@ppendix{A}{ }{#1}} 
\def\app{\@ppendix{A}{}{}} 
\def\subappendix#1#2{\global\advance\subno by 1\subsubno=0 
    \gdef\labeltype{\seclabel}%
    \ifssf\else\goodbreak\beforesubspace\fi 
    \global\ssffalse\nobreak 
    \noindent{\it #1\the\subno. #2\par}%
    \nobreak\subspace\noindent\ignorespaces} 
%
%
\def\@ck#1{\ifletter\bigskip\noindent\ignorespaces\else 
    \goodbreak\beforesecspace\nobreak 
    \noindent{\bf Acknowledgment#1\par}%
    \nobreak\secspace\noindent\ignorespaces\fi} 
\def\ack{\@ck{s}} 
\def\ackn{\@ck{}} 
\def\n@ip#1{\goodbreak\beforesecspace\nobreak 
    \noindent\smallfonts{\it #1}. \rm\ignorespaces} 
\def\naip{\n@ip{Note added in proof}} 
\def\na{\n@ip{Note added}} 
 
%
%
 
%
 
%
%
 
%
 
%
 
\def\tablecont{\topinsert\global\advance\tabno by -1 
    \tablecaption{(continued)}} 
\def\tablecaption#1{\gdef\labeltype{\tablabel}\global\widefalse 
    \leftskip=\secindent\parindent=0pt 
    \global\advance\tabno by 1 
    \smallfonts{\bf Table \ifappendix\applett\fi\the\tabno.} \rm #1\par 
    \smallskip\futurelet\next\t@b} 
\def\t@b{\ifx\next*\let\next=\widet@b 
             \else\ifx\next[\let\next=\fullwidet@b 
                      \else\let\next=\narrowt@b\fi\fi 
             \next} 
\def\widet@b#1{\global\widetrue\global\notfulltrue 
    \t@bwidth=\hsize\advance\t@bwidth by -\secindent}  
\def\fullwidet@b[#1]{\global\widetrue\global\notfullfalse 
    \leftskip=0pt\t@bwidth=\hsize}                   
\def\narrowt@b{\global\notfulltrue} 
\def\align{\catcode`?=13\ifnotfull\moveright\secindent\fi 
    \vbox\bgroup\halign\ifwide to \t@bwidth\fi 
    \bgroup\strut\tabskip=1.2pc plus1pc minus.5pc} 
\def\endalign{\egroup\egroup\catcode`?=12} 
 
%
%

%
%

%
 
%
%

%
 
\catcode`?=13 
\def\lineup{\setbox0=\hbox{\smallfonts\rm 0}%
    \digitwidth=\wd0%
    \def?{\kern\digitwidth}%
    \def\\{\hbox{$\phantom{-}$}}%
    \def\-{\llap{$-$}}} 
\catcode`?=12 
%
%
\def\sidetable#1#2{\hbox{\ifppt\hsize=18pc\t@bwidth=18pc 
                          \else\hsize=15pc\t@bwidth=15pc\fi 
    \parindent=0pt\vtop{\null #1\par}%
    \ifppt\hskip1.2pc\else\hskip1pc\fi 
    \vtop{\null #2\par}}}  
\def\lstable#1#2{\everypar{}\tempval=\hsize\hsize=\vsize 
    \vsize=\tempval\hoffset=-3pc 
    \global\tabno=#1\gdef\labeltype{\tablabel}%
    \noindent\smallfonts{\bf Table \ifappendix\applett\fi 
    \the\tabno.} \rm #2\par 
    \smallskip\futurelet\next\t@b} 
\def\inctabno{\global\advance\tabno by 1} 
%
%
 
%
 
%
\def\figure#1{\figc@ption{#1}\bigskip} 
\def\figc@ption#1{\global\advance\figno by 1\gdef\labeltype{\figlabel}%
   {\parindent=\secindent\smallfonts\hang 
    {\bf Figure \ifappendix\applett\fi\the\figno.} \rm #1\par}} 
%
%
\def\refHEAD{\goodbreak\beforesecspace 
     \noindent\textfonts{\bf References}\par 
     \let\ref=\rf 
     \nobreak\smallfonts\rm} 
\def\references{\refHEAD\parindent=0pt 
     \everypar{\hangindent=18pt\hangafter=1 
     \frenchspacing\rm}%
     \secspace} 
\def\rf#1{\par\noindent\hbox to 21pt{\hss #1\quad}\ignorespaces} 
%
 
%
 
%
%
\def\numrefjl#1#2#3#4#5{\par\rf{#1}#2 {\it #3 \bf #4} #5\par} 
%
%
\def\numrefbk#1#2#3#4{\par\rf{#1}#2 {\it #3} #4\par} 
%
%

%
%

%
\catcode`\@=12 
%
%
 
%
%
\def\jnlstyle{\pptfalse\headsize{14}{18}%
\textsize{10}{12}%
\smallsize{8}{10} 
\textind=16pt} 
%
%
 
%
%
 
%
\parindent=\textind 
%
\input epsf
\def\received#1{\insertspace 
     \parindent=\secindent\ifppt\textfonts\else\smallfonts\fi 
     \hang{Received #1}\rm } 
\def\figure#1{\global\advance\figno by 1\gdef\labeltype{\figlabel}%
   {\parindent=\secindent\smallfonts\hang 
    {\bf Figure \ifappendix\applett\fi\the\figno.} \rm #1\par}} 
\headline={\ifodd\pageno{\ifnum\pageno=\firstpage\titlehead
   \else\rrhead\fi}\else\lrhead\fi} 
\def\lpsn#1#2{LPSN-#1-LT#2}
 
\def\rrhead{\textfonts\hskip\secindent\it 
    \shorttitle\hfill\rm L\folio} 
\def\lrhead{\textfonts\hbox to\secindent{\rm L\folio\hss}%
    \it\aunames\hss} 
\footline={\ifnum\pageno=\firstpage{\smallfonts cond--mat/9408076}
\hfil\textfonts\rm L\folio\fi}   
\def\titlehead{\smallfonts J. Phys. A: Math. Gen.  {\bf 27} (1994) L579--L583 
\hfil\lpsn{94}{2}} 

\firstpage=579
\pageno=579

\jnlstyle
\jl{1}
\overfullrule=0pt

\letter{Fractal dimensions of confined clusters\hfill\break in two--dimensional 
directed percolation}[Letter to the Editor]

\author{C Kaiser and L Turban}[Letter to the Editor]
 
\address{Laboratoire de Physique du Solide\footnote\dag{Unit\'e de
Recherche Associ\'ee au CNRS No 155},  Universit\'e Henri Poincar\'e (Nancy~I),
BP 239, F--54506 Vand\oe uvre l\`es Nancy Cedex, France}

\received{15 June 1994}

\abs
The fractal structure of directed percolation
clusters, grown at the percolation threshold inside parabolic--like
systems, is studied in two dimensions via Monte--Carlo simulations. With a
free surface at $\scriptstyle y=\!\pm Cx^k$ and a dynamical exponent $\scriptstyle z$, the
surface shape is a relevant perturbation when $\scriptstyle k<1/z$ and the
fractal dimensions of the anisotropic clusters vary continuously with
$\scriptstyle k$. Analytic expressions for these variations are obtained using a
blob picture approach.
\endabs
\vglue1.5cm
\pacs{}
\submitted
\date

Since the work of Cardy [1], one knows that a large--scale
alteration of the shape of a system may influence its local critical
behaviour (further references can be found in the review~[2]). 

Cardy studied systems limited by corners or wedges which are scale invariant
shapes for an isotropic system and thus, introduce a marginal perturbation
leading to varying local exponents. More generally, one may consider systems
limited by a parabolic--like surface at $y\!=\!\pm Cx^k$ in the
$(x,y)$--plane in two dimensions, a parabolic cylinder or a paraboloid in
higher dimensions. Under a uniform rescaling of the lengths by a factor $b$,
$C$ is changed into $C'\!=\! b^{k-1}C$. As a consequence, $C$ grows to
infinity for $k\!>\!1$ whereas it goes to zero when $k\!<\!1$. It remains
scale invariant when $k\!=\!1$, which corresponds to the marginal corner
geometry mentioned above. 

Since $1/C$ vanishes for the the semi--infinite system, it can be
considered as a scaling field for the reference fixed point corresponding
to the flat surface. This scaling variable is an irrelevant perturbation
for $k\!>\!1$ and a relevant one for $k\!<\!1$. 

In the relevant case, the shape renormalizes to a narrow system, so that the
local order is weakened. Using a conformal transformation from the
half--space to the interior of the parabola, one can show that the critical
correlation functions decay from the tip along the $x$--axis in a stretched
exponential way~[3]. The order parameter at the tip also displays an
essential singularity as a function of the reduced temperature~[3, 4, 5].
Similar results were obtained for the percolation problem [6]. In the
case of chain polymers (self--avoiding--walks) with one end fixed at the
tip, the exponent associated with the radius of gyration becomes 
anisotropic~[7].

When the unperturbed system is itself anisotropic, i.e. when the
correlation length exponents~$\nu_\parallel$ in the $x$--direction and
$\nu_\perp$ in the transverse direction are different with a dynamical
exponent $z$ given by the ratio $\nu_\parallel/\nu_\perp$, the
appropriate scaling factor becomes $b^z$ in the $x$--direction. Then the
scaling field $1/C$ transforms as 
$$
{1\over C'}=b^{1-zk}\ {1\over C}\eqno(1)
$$
and the critical behaviour is modified when $k\leq 1/z$ instead of $1$. The
directed walk (polymer) with $z\!=\!2$ was studied in [8, 9]. In the
marginal case, which then corresponds to the true parabola, the local
susceptibilities given by the numbers of walks with $N$ steps
starting at the tip and ending either anywhere or on the surface,
diverge with a $C$--dependent power of $N$ as expected.

In the present work, we study the structure of directed percolation
clusters confined inside a "parabola". We consider the site percolation
problem, directed along the diagonal of a square lattice, with the sites
inside or on the "parabola" belonging to the system. 
{\par\begingroup\parindent=0pt\medskip
\epsfxsize=9truecm
\topinsert
\centerline{\epsfbox{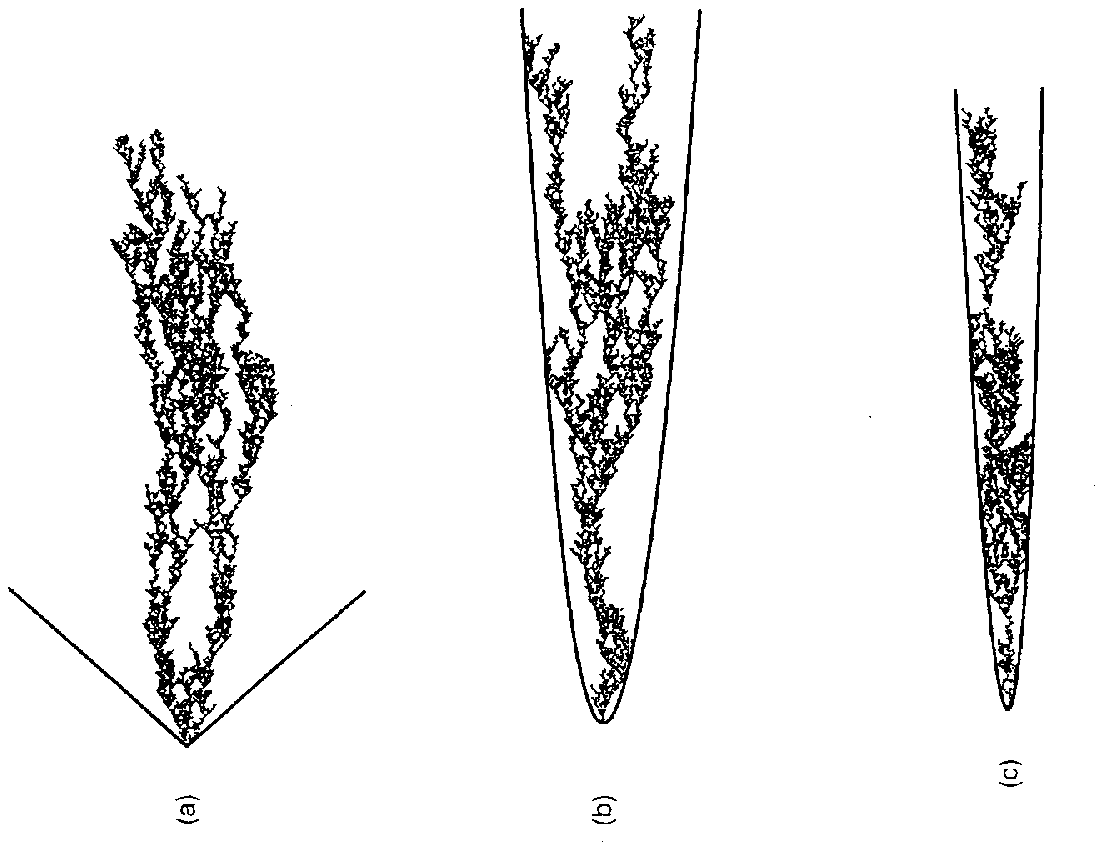}}
\smallskip
\figure{Structure of finite clusters generated at $p_c$
for: (a) free growth, (b--c) parabolic systems with $C\!=\!2,\ 4$ and
$k=1/2$.} 
\endinsert 
\endgroup
\par}
Using a standard Monte--Carlo method, $2.10^5$ clusters starting from the
tip at $x\!=\! y\!=\! 0$ were generated at the percolation threshold
$p_c\!=\! 0.705489(4)$ [10] as shown in figure 1. The square radii of
gyration for $s$--site clusters along the two directions
$$
X_s^2={1\over 2s^2}\sum_{i,j=1}^s(x_i-x_j)^2\qquad
Y_s^2={1\over 2s^2}\sum_{i,j=1}^s(y_i-y_j)^2\eqno(2)
$$
were put into bins with respect to the sizes $2^i\!\leq\! s\!<\!2^{i+1}$
($i\!=\!0,1,2,\dots$) and the bin averages $\langle X_s^2\rangle$, 
$\langle Y_s^2\rangle$, were calculated for "finite" clusters, i. e. for
clusters with $x_{\rm max}\!<\! L\!=\!1000$, the system size along the
$x$--direction.
{\par\begingroup\parindent=0pt\medskip
\epsfxsize=9truecm
\topinsert
\centerline{\epsfbox{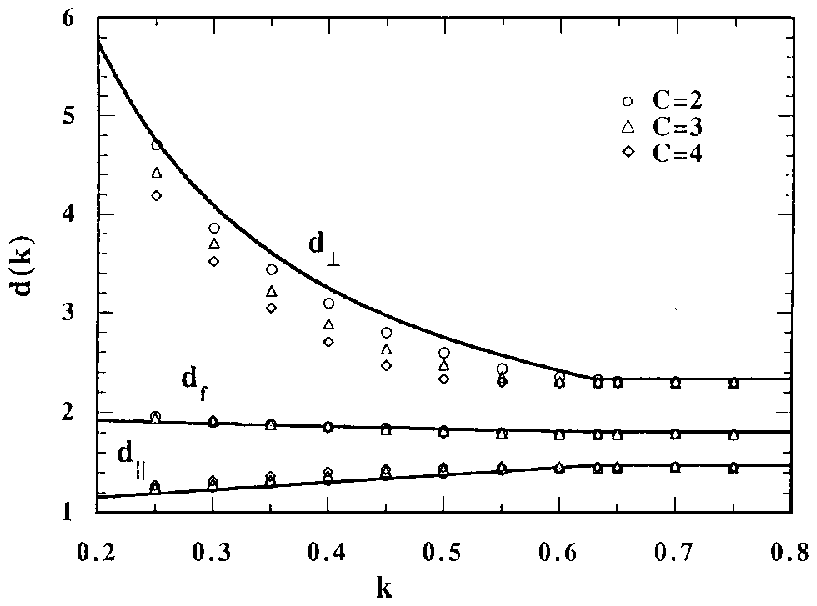}}
\smallskip
\figure{Fractal dimensions $d_\parallel(k)$,
$d_\perp(k)$ and $d_f(k)$ versus the system--shape exponent
$k$. The lines give the analytic results obtained via a blob picture
approach. The fractal dimensions are varying when the perturbation is
relevant $(k\!<\!1/z\!=\!0.633(2))$.}
\endinsert 
\endgroup
\par}
Asymptotically, the following power laws are expected
$$
\overline{s}\sim\overline{X_s}^{\,d_\parallel(k)}
\sim\overline{Y_s}^{\,d_\perp(k)}\eqno(3)
$$
where $\overline{s}$ is the center of the bin, 
$\overline{X_s}\!=\!\langle X_s^2\rangle^{1/2}$, 
$\overline{Y_s}\!=\!\langle Y_s^2\rangle^{1/2}$ and ${d_\parallel}(k)$,
${d_\perp}(k)$, are the fractal dimensions in the two directions. In a
log--log plot, a linear behaviour is obtained for
$2^6\!<\overline{s}\!<\!2^{12}$--$\,2^{13}$, the deviations at larger sizes
being mainly due to the finite size of the system. There, the contribution
from longer clusters with the same values of $\overline{s}$ is not taken
into account. The fractal dimensions, shown in figure 2 versus $k$ for
different values of $C$, were obtained from least--square fits of the data
in the intermediate linear regime.

In order to analyse our results, let us first consider unconfined
directed percolation [11]. In two dimensions, the best estimates for the
critical exponents, obtained through series expansions, are [10]:
$$
\nu_\perp=1.097(2)\quad\nu_\parallel=1.734(2)\quad\gamma=2.278(2)
\eqno(4)
$$
giving
$$
z=1.581(3)\qquad\beta=0.276(3)
\eqno(5)
$$
from scaling. Extending an argument of Stauffer for ordinary 
percolation~[12], one obtains the following relations for the fractal
dimensions of critical directed percolation clusters:
$$
{d_\parallel}={\beta+\gamma\over
\nu_\parallel}= 1.473(2)\qquad{d_\perp}={\beta+\gamma\over\nu_\perp}=2.329(3).
\eqno(6)
$$
The numerical value of ${d_\perp}\!>\! d$ may be at first sight surprising
for a fractal object. Actually, it can be traced to the anisotropy. Let
${l_\parallel}$ and ${l_\perp}$ be the length and width of a $s$--site
cluster with $s\sim{l_\parallel}^{d_\parallel}\sim{l_\perp}^{d_\perp}$. It
follows that $$
{l_\parallel}={l_\perp}^{\,{d_\perp}/{d_\parallel}}={l_\perp}^z\eqno(7)
$$
grows faster with $s$ than ${l_\perp}$. $s$ itself grows faster with
${l_\perp}$ than an isotropic massive object due to the parallel
growth. Introducing the characteristic length associated with the surface
of the cluster 
$$
l\sim({l_\parallel}{l_\perp})^{1/2}\sim s^{1/d_f}
\eqno(8)
$$
a single fractal dimension $d_f$ can be defined with [11]
$$
{1\over d_f}={1\over 2}\left({1\over{d_\parallel}}+{1\over{d_\perp}}\right)\eqno(9)
$$
and now, $d_f\!=\!1.805(2)\!<\!2$. 
{\par\begingroup\parindent=0pt\medskip
\epsfxsize=9truecm
\topinsert
\centerline{\epsfbox{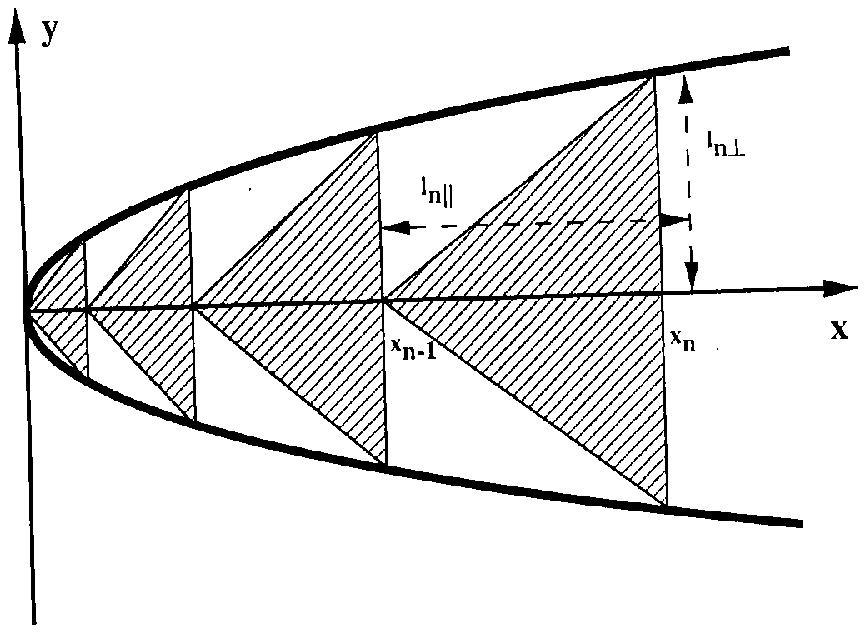}}
\smallskip
\figure{In the blob picture approach, confined clusters are built
through piling up successive anisotropic blobs, with a transverse size
$l_{n\perp}$ fixed by the geometry of the system. The correlations
inside each blob are the same as for unconfined clusters.}
\endinsert 
\endgroup
\par}
When the perturbation introduced by the surface is relevant, the structure
of confined clusters can be studied studied using a blob picture approach
[13, 14, 7]. Inside the "parabola", the cluster configuration results from
the piling up of anisotropic blobs as shown in figure 3. Within each blob,
the correlations are the same as for unconfined clusters. The $n$th~blob
at~$x_n$ has a width~${l_{n\perp}}\!=\!Cx_n^k$ which is fixed by the
geometry of the system. Using equation~(7), its length is 
$$
{l_{n\parallel}}=C^zx_n^{zk}\sim C^un^v\eqno(10)
$$
where the last expression corresponds to a continuum approximation in the
blob index~$n$. {}From $x_n\!-\!~x_{n-1}\!=\!{l_{n\parallel}}$, one deduces the
following differential equation in the continuum limit,
$$
{\d x_n\over\d n}\sim C^un^v\sim C^{(u-z)/zk}n^{-1+v/zk}\eqno(11)
$$
where the last term follows from the $n$--dependance of $x_n$ in~(10).
Identifying powers of~$C$ and~$n$, one obtains $u\!=\!~z/(1-zk)$ and
$v\!=\! ku$. These exponents are positive for a relevant perturbation
only. Otherwise, when $k\!>\!1/z$, the cluster grows freely, i. e. there
is a single blob. Let $s_n\!\sim\!({l_{n\parallel}})^{d_\parallel}$
be the number of sites inside the $n$th blob. For the whole cluster, we
have: 
$$ 
\fl s=\sum_{n=1}^Ns_n\sim C^{z{d_\parallel}/(1-zk)}\!\int_1^N\!\!\d n\ 
n^{zk{d_\parallel}/(1-zk)}\sim C^{z{d_\parallel}/(1-zk)}\  N^{[1+zk({d_\parallel}-1)]/(1-zk)}.
\eqno(12)
$$
In the same way, the cluster length is given by
$$
{l_\parallel}=\sum_{n=1}^N{l_{n\parallel}}\sim C^{z/(1-zk)}\!\int_1^N\!\!\d n\  n^{zk/(1-zk)}\sim
C^{z/(1-zk)}\  N^{1/(1-zk)}.\eqno(13)
$$
Eliminating $N$ between (12) and (13), one finally obtains:
$$
\fl s\sim C^{z({d_\parallel}-1)}\  {l_\parallel}^{d_\parallel(k)}\qquad
d_\parallel(k)=1+zk({d_\parallel}-1)\qquad0\leq k\leq 1/z
\eqno(14) 
$$ 
which, together with
${l_\perp}\!=\! C{l_\parallel}^k$, also gives: 
$$
\fl s\sim C^{-1/k}\  {l_\perp}^{d_\perp(k)}\qquad
d_\perp(k)={d_\parallel(k)\over k}\qquad0<k\leq1/z .
\eqno(15) 
$$

The variations of the fractal dimensions are shown in figure~2. For a
relevant perturbation, $d_\parallel(k)$ and $d_\perp(k)$ vary continuously
with $k$ until they reach their unperturbed values for $k\!=\!1/z$. As
a result of finite--size effects, the  Monte--Carlo values deviate
from the analytical ones when $C$ increases. For large $C$ values the
system is more open and the number of blobs inside it remains small for a
finite system. As a consequence, the Monte--Carlo estimates for the fractal
dimensions are then closer to the unperturbed ones. 

Results for the tip percolation probability will be discussed elsewhere.

\ack
CK thanks Henri Poincar\'e University for hospitality. This work was
supported by CNIMAT under project No 155/93. 

\references
\numrefjl{[1]}{Cardy J L 1983}{\JPA}{16}{3617}
\numrefjl{[2]}{Igl\'oi F, Peschel I and Turban L 1993}{Adv. Phys.}{42}{683}
\numrefjl{[3]}{Peschel I, Turban L and Igl\'oi F 1991}{\JPA}{24}{L1229}
\numrefjl{[4]}{Davies B and Peschel I 1992}{Ann. Phys. (Leipzig)}{2}{79}
\numrefjl{[5]}{Blawid S and Peschel I 1994}{\ZP\ \rm B}{95}{73}
\numrefjl{[6]}{Berche B, Debierre J M and Eckle P}{1994}{}{Nancy I
preprint}  
\numrefjl{[7]}{Turban L and Berche B 1993}{\JP\  I}{3}{925}
\numrefjl{[8]}{Turban L 1992}{\JPA}{25}{L127} 
\numrefjl{[9]}{Igl\'oi F 1992}{\PR\ \rm B}{45}{7024} 
\numrefjl{[10]}{Essam J W, Guttmann A J and De'Bell K 1988}{\JPA}{21}{3815}
\numrefjl{[11]}{Kinzel W 1983}{Ann. Israel Phys. Soc.}{5}{425}
\numrefbk{[12]}{Stauffer D 1985}{Introduction to Percolation
Theory}{(London: Taylor \& Francis) p 64}
\numrefjl{[13]}{Pincus P 1976}{Macromolecules}{9}{386}
\numrefbk{[14]}{de Gennes P G 1979}{Scaling Concepts in Polymer
Physics}{(Ithaca: Cornell University Press) p~50}

\vfill\eject\bye